\documentclass[preprintnumbers]{revtex4}
\usepackage{amssymb}
\usepackage{amsmath}
\usepackage{graphicx}
\usepackage{dcolumn}
\usepackage{bm}
\usepackage{subfigure}
\usepackage{color}

\begin{document}

\title{Hawking-Page Phase Transition of the four-dimensional de-Sitter Spacetime with non-linear source}
\author{Yun-Zhi Du,$^{1,2}$ Huai-Fan Li,$^{1,2}$ Li-Chun Zhang$^{1,2,*}$}
\address{$^1$Department of Physics, Shanxi Datong University, Datong 037009, China\\
$^2$Institute of Theoretical Physics, Shanxi Datong University, Datong, 037009, China}

\thanks{\emph{e-mail:zhlc2969@163.com(Corresponding author)}}
\thanks{\emph{duyzh13@lzu.edu.cn}}
\thanks{\emph{huaifan999@163.com}}

\begin{abstract}
The interplay between a dS black hole and cosmological horizons introduces distinctive thermodynamic behavior in a dS spacetime (for example the well-known upper bounds of mass and entropy in Class. Quant. Grav. 37 (2020) 5). Based on this point, we present the Hawking-Page (HP) phase transition of the four-dimensional dS spacetime with non-linear charge correction when the effective pressure is fixed, and analyze the effects of different effective pressures and non-linear charge corrections on HP phase transition. The evolution of this dS spacetime undergoing the HP phase transition is also investigated. We find that the existent curve of HP phase transition is a closed one with two different branches. That means there exist the upper bounds of the HP temperature and HP pressure, which is completely distinguished with that in AdS spacetime. And with the decreasing of the distance between two horizons, the dS spacetime at the coexistent curve of HP phase transition is going along with different branches. Furthermore we also explore the influence of charge and non-linear charge correction on the coexistent curve.
\end{abstract}

\maketitle

\section{Introduction}

Since the discovery of the Hawking-Page (HP) phase transition \cite{Hawking1983}, the phase transition in a spacetime with an/a AdS/dS black hole has been widely investigated in the extended phase space \cite{Dinsmore2020,Cai1306,Wei1209,Caldarelli2000,Mann1207,Wei2015,Banerjee1109,Hendi1702,Bhattacharya2017,Zeng2017,Hendi1803,Zhang1502,Cheng1603,
Zou1702,Dolan2014,Altamirano2013,Altamirano2014,Du2021,Zhang2020}. The HP phase transition of an AdS spacetime gave the evolution of spacetime with different phases. Namely, for an AdS spacetime with the increasing of temperature the dominant configuration is from the pure thermal radiation phase, then to the coexistent phase with an AdS black hole and thermal radiation, and finally to a stable black hole. This HP phase transition was explained by Witten \cite{Witten1998} as a confinement/deconfinement phase transition in gauge theory. And it could also be understood as a solid/liquid phase transition \cite{Altamirano2013} by regarding the cosmological constant as pressure $P=-\frac{\Lambda}{8\pi}=\frac{(n-1)(n-2)}{16\pi l^2}$, whose conjugate variable is the thermodynamic volume.

For the Schwarzschild-AdS black holes with hyperbolic horizons, which are the thermal stable, the HP phase transition does not emerge. While for Schwarzschild-AdS black holes with spherical horizon, there exist the HP phase transition between the pure thermal radiation in AdS spacetime and stable larger black holes. Subsequently the authors in \cite{Chamblin1999} had extended to the charge AdS (i.e., Reissner-Nordstrom-AdS) black hole. The HP phase transition in the Einstein-Gauss-Bonnet gravity also was investigated \cite{Wei2020,Wang2021}. Currently, there are some researches on the HP phase transition in an AdS spacetime. Therefore it is a natural question whether HP phase transition can survive in a dS spacetime. In Ref. \cite{Zhao2020,Su2021} the authors revealed the relationship of HP phase transition properties in dS black holes and their specific boundary in different extended phase spaces by putting black holes into a spherical cavity. However the boundary of dS black hole is artificially added, which will lead to lose its universality. Based on this issue we will investigate the HP phase transition of a dS spacetime with non-linear source through considering the interplay between dS black hole horizon and cosmological horizons.

In nature, most physical systems are non-linear, so the non-linear field theories are of interest to different branches of mathematical physics. The non-linear electrodynamics (NLED) have the richer structures and they can reduce to linear Maxwell theory (LMT) in special case. Since in LMT there are various limitations \cite{Heisenberg1936,Yajima2001,Schwinger1951}, especially the limitation about the radiation propagation inside specific materials \cite{Lorenci2001,Lorenci2002,Novello2003,Novello2012}, NLED should be considered more. In addition NLED objects can remove both of the big bang and black hole singularities \cite{Cavaglia2003}. Recently the authors \cite{Dehghani2005,Du2021,Zhang2020} checked the first law of thermodynamics for the $n+1$-dimensional topological static black hole with the mentioned NLED and analyzed the effect of the non-linear charge correction on the thermodynamic properties of black hole. In this work we will present the corresponding properties of HP phase transition in this spacetime. And a unique phenomena will be exhibited in coexistent curve of HP phase transition temperature and HP phase transition pressure.

This work is organized as follows: in section \ref{two}, we present the thermodynamic quantities of the four-dimensional dS spacetime with non-linear source and analyze the effect of different non-linear charge corrections and different effective pressures on thermodynamic quantities. Then in section \ref{three}, we investigate the Gibbs free energy to explore the evolution of this dS spacetime and discuss the property of HP phase transition. Furthermore the coexistent curve of HP phase transition temperature and HP phase transition pressure is presented. We also analyze the influence of the non-linear charge correction on HP phase transition. Finally, a brief summary is given in Sec. \ref{four}.

\section{The thermodynamical quantities of the four-dimensional de-Sitter spacetime with non-linear source}
\label{two}

The static spherically symmetric black hole solution in a four-dimensional spacetime with non-linear source were given as \cite{Du2021,Zhang2020,Hendi2015,Dehghani2005}
\begin{eqnarray}
d s^{2}&=&-f(r) d t^{2}+f^{-1} d r^{2}+r^{2} d \Omega_{2}^{2},\\
f(r)&=&k-\frac{2 M}{r}-\frac{\Lambda r^2}{3}+\frac{q^2}{r^2}-\frac{2q^4\alpha}{5r^6}.\label{f}
\end{eqnarray}
Here $M$ and $q$ are the black hole mass and charge, and the last term in eq. (\ref{f}) indicates the effect of the non-linearity. For similarity, we redefined the non-linear source term as $\bar{\phi}=\frac{q^4\alpha}{r_+^6}$ and recall $\bar\phi$ as the non-linear charge correction. In the following we mainly focus on the the solution $\Lambda>0$ with $k=1$, i.e., the de-Sitter spacetime with a black hole. In this system there are two horizons, one is of dS black hole ($r_+$), another is of cosmology ($r_c$). And these two horizons are satisfied with the expression $f(r_{+,c})=0$. The radiation temperatures at two horizons were given in Refs. \cite{Heisenberg1936,Hendi2015}. When regarding the four-dimensional dS spacetime with non-linear source as an ordinary thermodynamic system in the thermodynamic equilibrium and considering the correlations of the two horizons, the effective thermodynamic quantities ($T_{eff},~P_{eff},~V,~S,~\Phi_{eff}$) can be calculated. Here we point out the entropy is not only the sum of two horizons, it also contains the connection between two horizons. Considering the connection between the black hole horizon and the cosmological horizon, the corresponding first law of black hole thermodynamics is given by \cite{Zhang2019}
\begin{equation}
dM=T_{eff}dS-P_{eff}dV+\Phi_{eff}dq.
\end{equation}
Note that the thermodynamic volume is that between the black hole horizon $r_+$ and the cosmological horizon $r_c$ \cite{Ali2019}
\begin{eqnarray}
V=\frac{V_2r_+^3(1-x^3)}{3x^2},~~~~~~S=\frac{V_2r_+^2F(x)}{4x^2} \label{SV}
\end{eqnarray}
with $V_2=\frac{2\pi^{3/2}}{\Gamma(3/2)}$, $x=\frac{r_+}{r_c}$, and
\begin{eqnarray}
F(x)=\frac{8}{5}(1-x^3)^{\frac{2}{3}}-\frac{2(4-5x^3-x^5)}{5(1-x^3)}+1+x^2
=\bar f(x)+1+x^2.
\end{eqnarray}
Note that the total entropy is not only the sum of entropy at the black hole horizon and the cosmological horizon, and $\bar f(x)$ in the form of $F(x)$ represents the extra contribution from the correlations of the two horizons.

The effective temperature, effective pressure, and mass were shown as the following in Ref. \cite{Zhang2020,Du2021}
\begin{eqnarray}
T_{eff}&=&\frac{f_2(x)}{f_1(x)r_+}-\frac{q^2f_3(x)}{f_1(x)r_+^3},\label{T}\\
P_{eff}&=&-\frac{f_5(x)}{f_4(x)r_+^2}-\frac{q^2f_6(x)}{f_4(x)r_+^4}\label{P},\\
M&=&\frac{V_2r_+(1-x^2)}{8\pi(1-x^3)}\left[k+\frac{q^2(1+x^2)}{r_+^2}-\frac{2\bar\phi}{5}\right]\label{MM}
\end{eqnarray}
with
\begin{eqnarray}
f_1(x)\!&=&\!\frac{4\pi(1+x^4)}{1-x},~~f_3(x)=(1+x+x^2)(1+x^4)-2x^3,~~f_4(x)=\frac{8\pi(1+x^4)}{x(1-x)},\nonumber\\
f_2(x)\!&=&\!\frac{2\bar\phi}{5}\left[5(1+x+x^2)(1+x^8)+2x^3(1+x+x^2+x^3+x^4)\right]\nonumber\\
&&+k\left[(1-3x^2)(1+x+x^2)+4x^3(1+x)\right],\nonumber\\
f_5(x)\!&=&\!kx(1+x)F'/2-\frac{k(1+2x)F}{1+x+x^2}-\bar\phi F'(1+x)(1+x^2)(1+x^4)\nonumber\\
&&-\frac{2\bar\phi F(5+10x+15x^2+12x^3+9x^4+6x^5+3x^6)}{5(1+x+x^2)},\nonumber\\
f_6(x)\!&=&\!\frac{(1+2x+3x^2)F}{1+x+x^2}+x(1+x)(1+x^2)F'/2.\nonumber
\end{eqnarray}

For this thermodynamic system undergoing an isobaric process, from eq. (\ref{P}) we find that the physical horizon radius $r_+$ satisfies the following form
\begin{eqnarray}
r_+=r_p=\sqrt{\frac{-f_5(x)+\sqrt{f_5^2(x)-4q^2P_{eff}f_4(x)f_6(x)}}{2P_{eff}f_4(x)}}.\label{rp}
\end{eqnarray}

\begin{figure}[htp]
\subfigure[$~~\bar\phi=0.002$]{\includegraphics[width=0.35\textwidth]{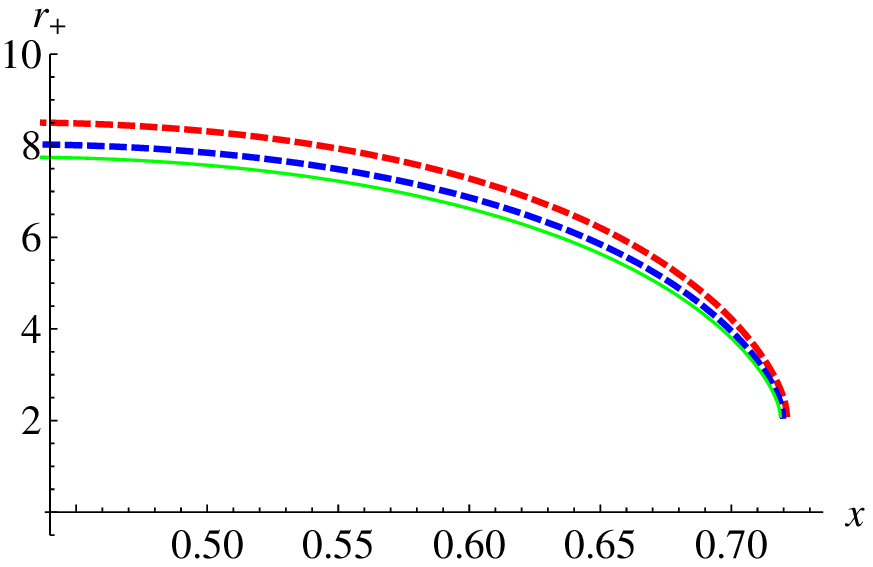}\label{rxp}}~~~~
\subfigure[$~~P_{eff}=1.459\times 10^{-4}$]{\includegraphics[width=0.35\textwidth]{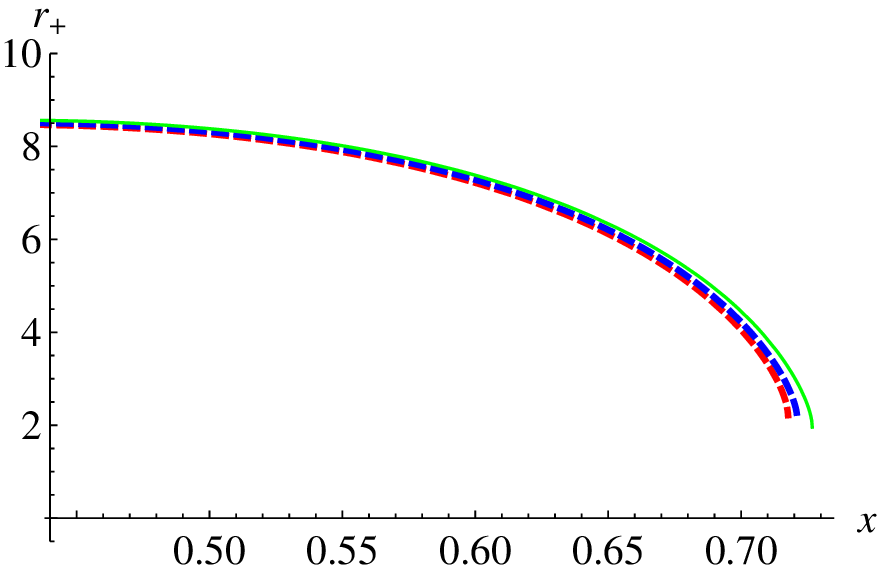}\label{rxphi}}
\caption{The horizon of black hole $r_+$ is as a function of $x$ for the parameters $k=1,~q=1$. In the left picture the effective pressure set to $P_{eff}=1.459\times 10^{-4}$ (dashed red thick line), $P_{eff}=1.634\times 10^{-4}$ (dashed blue thick line), $P_{eff}=1.751\times 10^{-4}$ (thin green line), respectively. In the right picture the non-linear charge correction set to  $\bar\phi=0$ (dashed red thick line), $\bar\phi=0.002$ (dashed blue thick line), $\bar\phi=0.005$ (thin green line), respectively.}\label{rx}
\end{figure}

In the following we mainly focus on the thermodynamic properties of the four-dimensional dS spacetime with the non-linear source undergoing the isobaric process. With eq. (\ref{rp}) the horizon of black hole with $x$ for different effective pressures and non-linear charge corrections are exhibited in Fig. \ref{rx}. It is obviously that for the dS spacetime with the fixed non-linear charge correction and effective pressure, the radius of black hole horizon is decreasing monotonously with $x$ ($0.44\leq x\leq0.73$). And it is decreasing with the effective pressure, and increasing with the non-linear charge correction.

\begin{figure}[htp]
\subfigure[$~~\bar\phi=0.002$]{\includegraphics[width=0.35\textwidth]{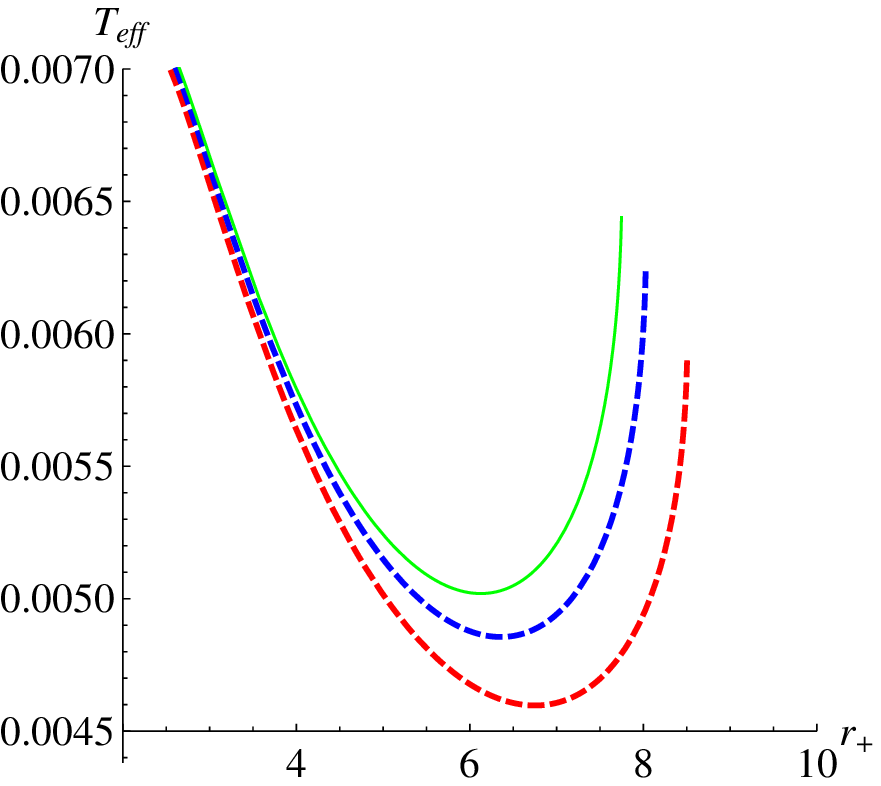}\label{trp}}~~~
\subfigure[$~~\bar\phi=0.002$]{\includegraphics[width=0.35\textwidth]{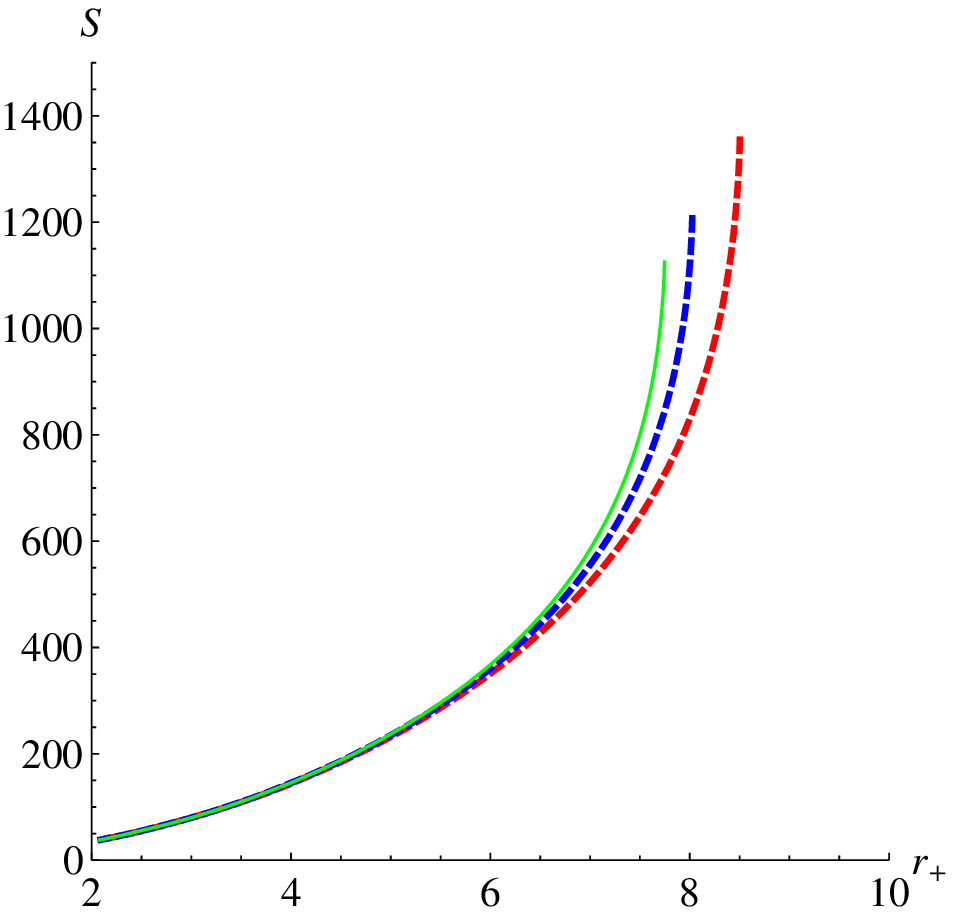}\label{srp}}~~~
\caption{For the given non-linear charge correction, the effective temperature $T_{eff}$ and entropy $S$ are both as a function of $r_+$ for the parameters $k=1,~q=1$. The effective pressure $P_{eff}=1.459\times 10^{-4}$ (dashed red thick lines), $1.634\times 10^{-4}$ (dashed blue thick lines), $1.751\times 10^{-4}$ (thin green lines), respectively.}\label{tsrp}
\end{figure}

\begin{figure}[htp]
\subfigure[$~~P_{eff}=1.459\times 10^{-4}$]{\includegraphics[width=0.35\textwidth]{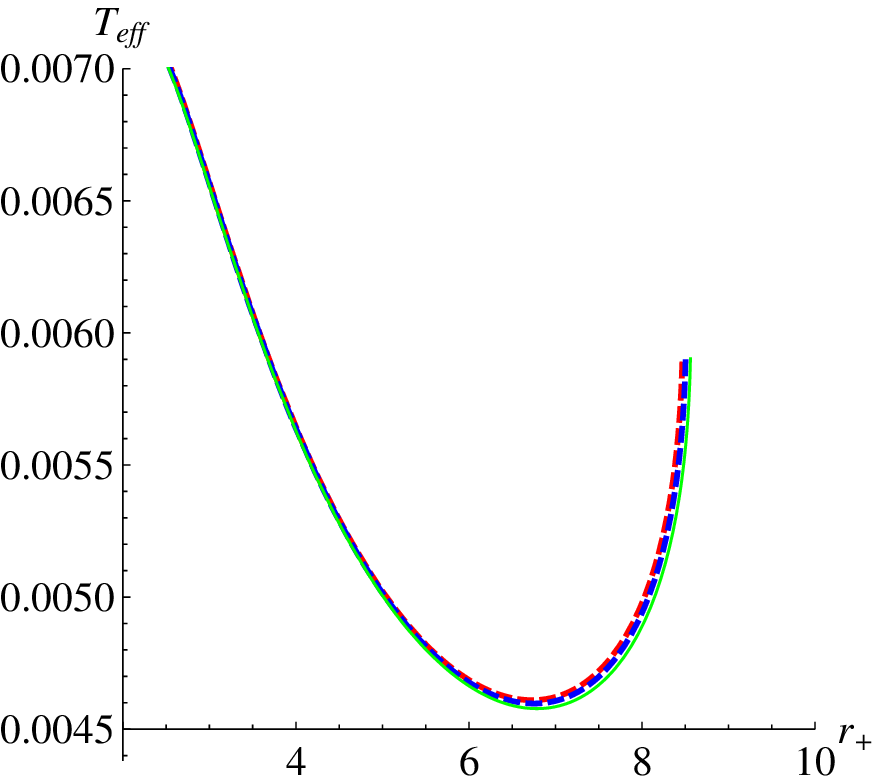}\label{trphi}}~~~
\subfigure[$~~P_{eff}=1.459\times 10^{-4}$]{\includegraphics[width=0.35\textwidth]{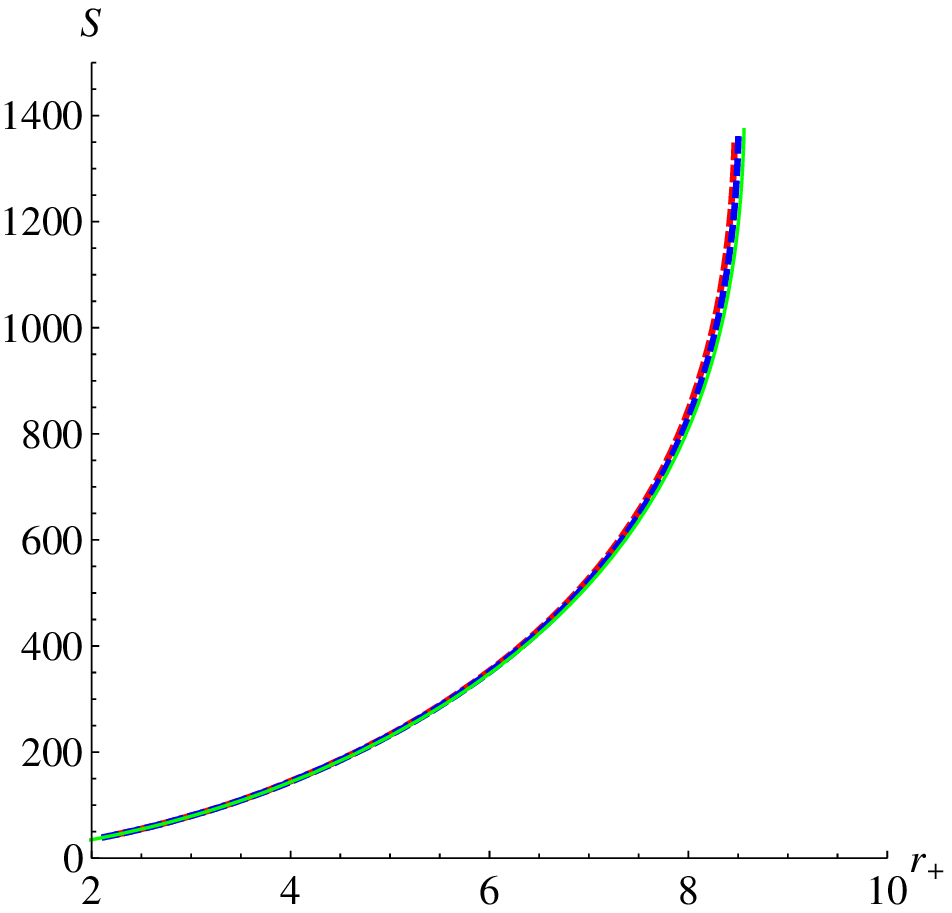}\label{srphi}}~~~
\caption{For the given effective pressure, the effective temperature $T_{eff}$ and entropy $S$ are both as a function of $r_+$ for the parameters $k=1,~q=1$. The non-linear charge correction $\bar\phi=0$ (dashed red thick lines), $0.002$ (dashed blue thick lines), $0.005$ (thin green lines), respectively.\label{tsrphi}}
\end{figure}

For the given non-linear charge correction and the given effective pressure, the effective temperature and entropy with $r_+$ are showed in Figs. \ref{tsrp} and \ref{tsrphi}, respectively. From Figs. \ref{trp} and \ref{trphi}, we know that for the fixed horizon $r_+$ the effective temperature increases with the effective pressure and decreases with the non-linear charge correction. Furthermore it decreases with the lower horizon radius for fixed effective pressure and increases with the bigger horizon radius for fixed non-linear charge correction. Note that there exists the minimal effective temperature $T_{eff}^0$ for the dS black hole with fixed effective pressure and non-linear charge correction. As $T_{eff}<T_{eff}^0$, there is no dS black hole, while there exists a pair of dS black holes when $T_{eff}>T_{eff}^0$. While from Figs. \ref{srp} and \ref{srphi}, for the given effective pressure and nonlinear charge correction the entropy increases monotonously with $r_+$. And for the fixed horizon $r_+$ it increases with the effective pressure and decreases with the non-linear charge correction. These properties are similar to that of the effective temperature. Especially for $T_{eff}>T_{eff}^0$, the heat capacity at constant pressure is negative for the smaller black holes, while it is positive for the bigger black holes. That means in dS spacetime with $T_{eff}>T_{eff}^0$ there exist the stable bigger black holes, instead of the smaller black holes.

\section{Hawking-Page Phase Transition of the dS spacetime with non-linear source}
\label{three}
As is well know, the horizons of black hole and cosmology are both of the Hawking radiations. These Hawking radiations can be regarded the background heat bath. When the dS spacetime is in the thermodynamic equilibrium, the dS black hole can be regarded in the background heat bath. Here the energy can be exchanged between the dS black hole and the background heat bath and the Gibbs free energy should be zero. Gibbs free energy is an important thermodynamic quantity to investigate the phase transition in addition to the equal area law. It can be used to study the HP phase transition, i.e., the HP phase transition emerges at $G=0$. In this part we will analyze the properties of HP phase transition in the four-dimensional dS spacetime with non-linear source undergoing an isobaric process and give the curve of $T_{eff}^{HP}-P_{eff}^{HP}$.

\begin{figure}[htp]
\subfigure[$~~\bar\phi=0.002$]{\includegraphics[width=0.4\textwidth]{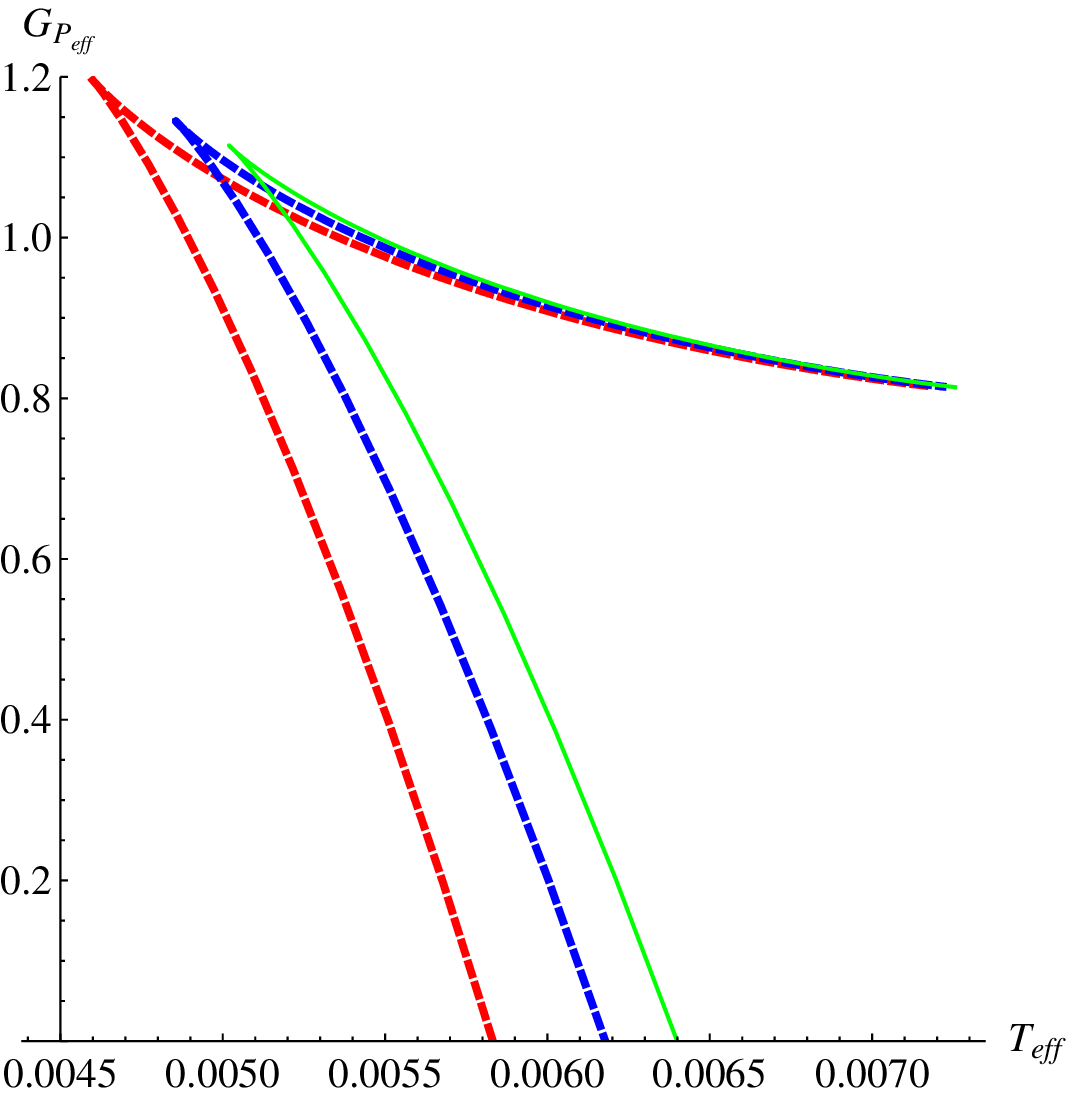}\label{gtp}}~~~
\subfigure[$~~P_{eff}=1.459\times10^{-4}$]{\includegraphics[width=0.4\textwidth]{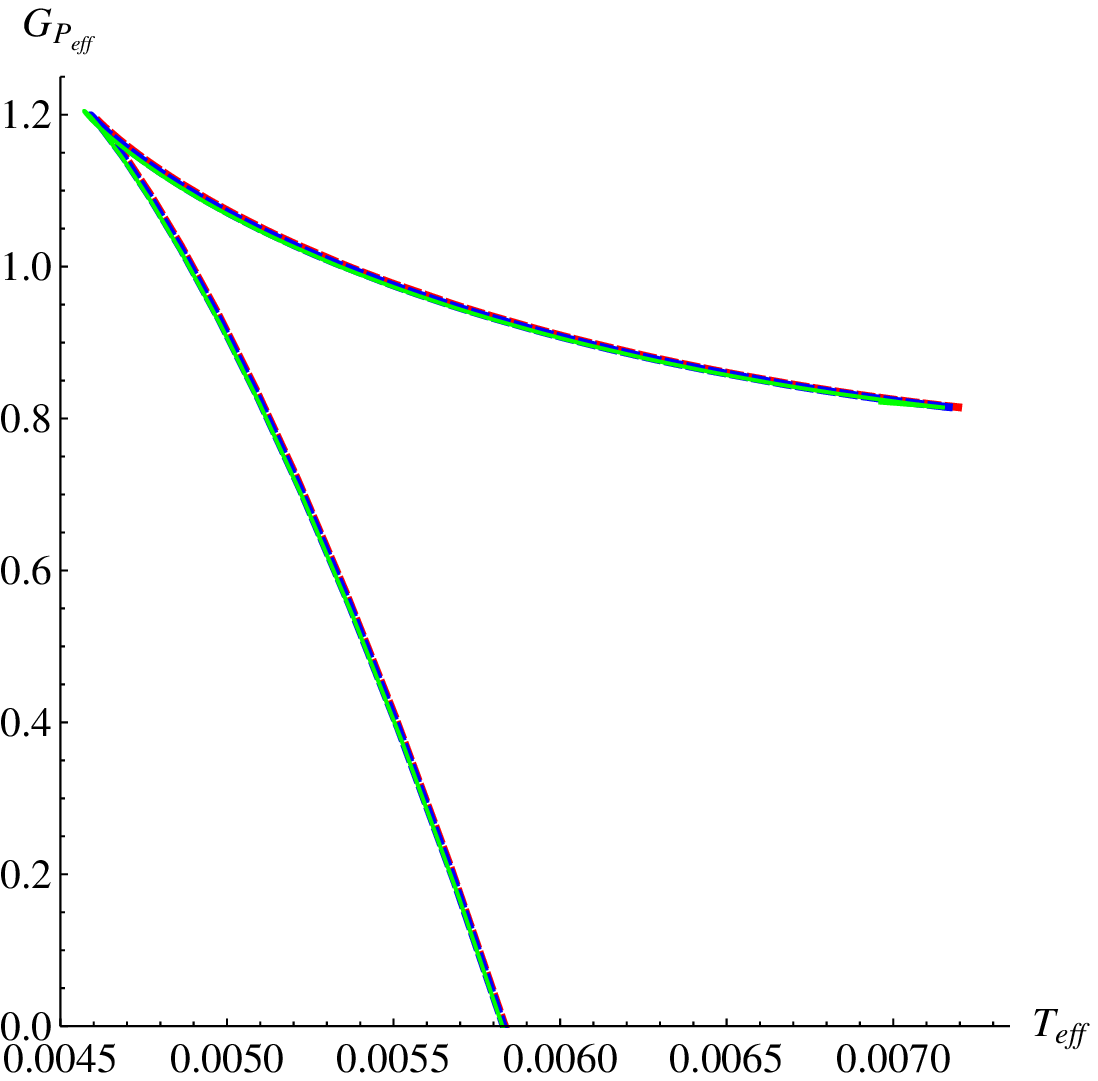}\label{gtphi}}~~~
\caption{The Gibbs free energy $G_{P_{eff}}$ as a function of $T_{eff}$ with the parameters $k=1,~q=1$ for different effective pressures and non-linear charge corrections. In the left, The effective pressure $P_{eff}=1.459\times 10^{-4}$ (dashed red thick lines), $1.634\times 10^{-4}$ (dashed blue thick lines), $1.751\times 10^{-4}$ (green thin lines), respectively. In the right, the non-linear charge correction $\bar\phi=0$ (dashed red thick lines), $0.002$ (dashed blue thick lines), $0.005$ (green thin lines), respectively.\label{gt}}
\end{figure}

In the four-dimensional dS spacetime with non-linear source, the Gibbs free energy reads
\begin{eqnarray}
G(r_+,x)=M-T_{eff}S+P_{eff}V.
\end{eqnarray}
With eqs. (\ref{SV}), (\ref{T}), (\ref{P}), (\ref{MM}), and (\ref{rp}), we have shown the pictures of $G$ with $q=1,~k=1$ for the given effective pressure and the given non-linear charge correction in Fig. \ref{gt}. It is obviously that the Gibbs free energy is not the monotonic function of $T_{eff}$, there are two branches corresponding to small and large dS black hole phases. And the Gibbs free energys of these two dS black holes are obviously increasing with the effective pressure when the effective pressure is fixed. While the Gibbs free energy of two different dS black holes are both slightly decreasing with the non-linear charge correction. It is clear that in Fig. \ref{gt} there exist the inflexion points and intersections with the given effective pressures and non-linear charge corrections. That means there exist the minimum effective temperature $T_{eff}^0$ (the inflexion point) and the HP effective temperature $T_{eff}^{HP}$ (the intersection), $T_{eff}^0<T_{eff}^{HP}$. Furthermore, $T_{eff}^0$ and $T_{eff}^{HP}$ are both increasing with the effective pressure, while they are slightly deceasing with the non-linear charge correction.

\begin{figure}[htp]
\subfigure[$~~\bar\phi=0.002,~P_{eff}=1.459\times 10^{-4}$]{\includegraphics[width=0.4\textwidth]{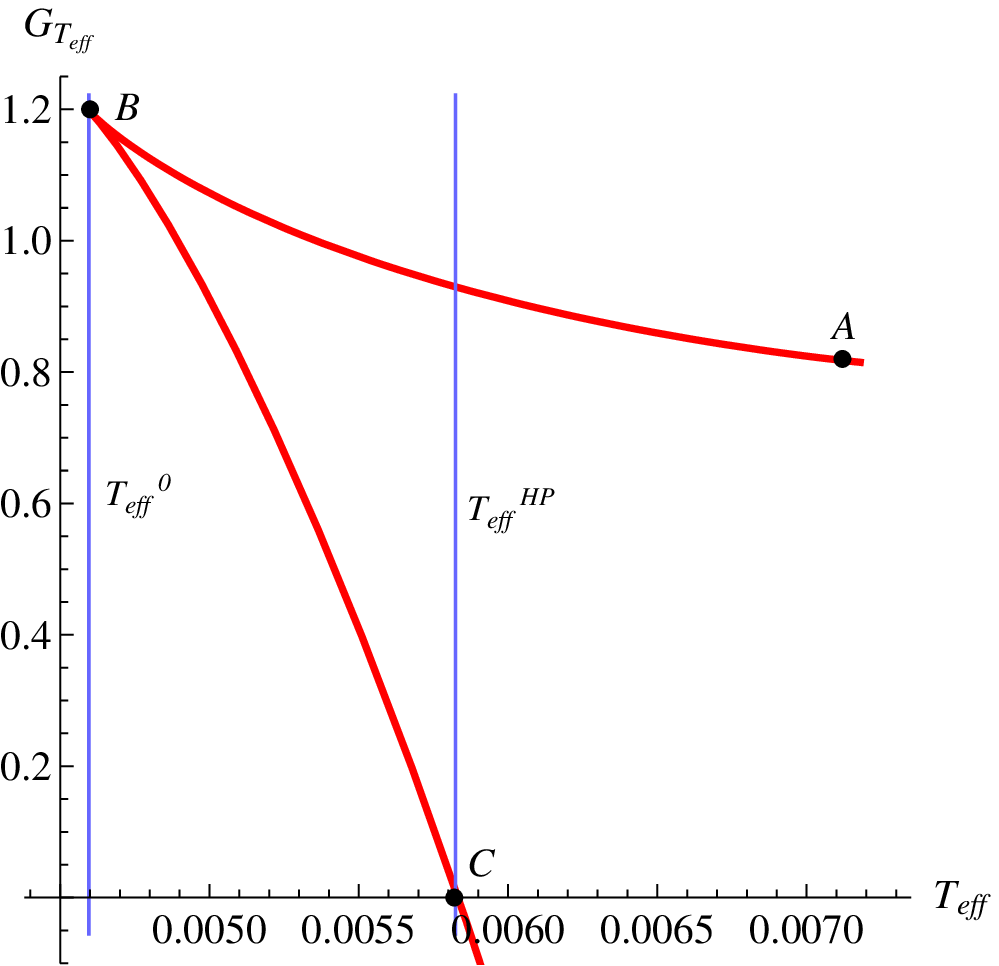}\label{gt25}}~~~
\subfigure[$~~\bar\phi=0.002,~P_{eff}=1.459\times 10^{-4}$]{\includegraphics[width=0.4\textwidth]{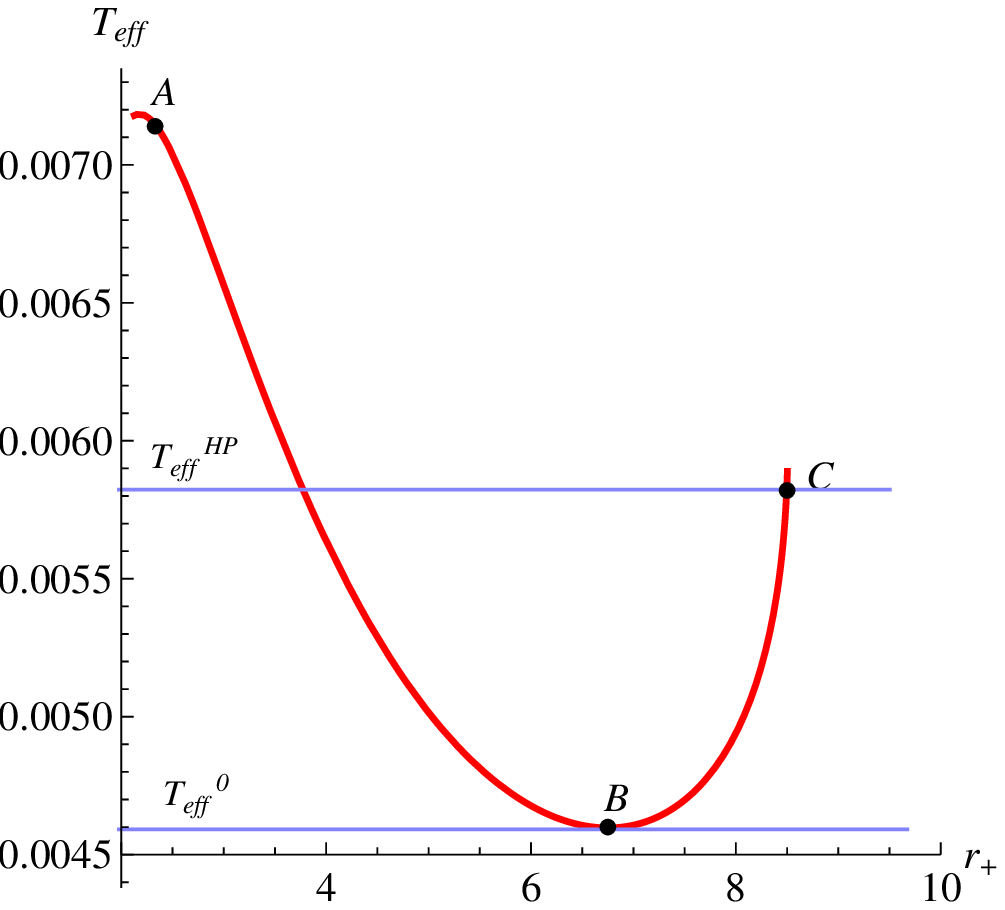}\label{tr25}}~~~
\caption{The pictures of $G_{P_{eff}}-T_{eff}$ and $T_{eff}-r_+$ for the parameters $k=1,~q=1$. \label{gtsr25}}
\end{figure}

\begin{figure}[htp]
\subfigure[$~~\bar\phi=0.002,~P_{eff}=1.459\times 10^{-4}$]{\includegraphics[width=0.35\textwidth]{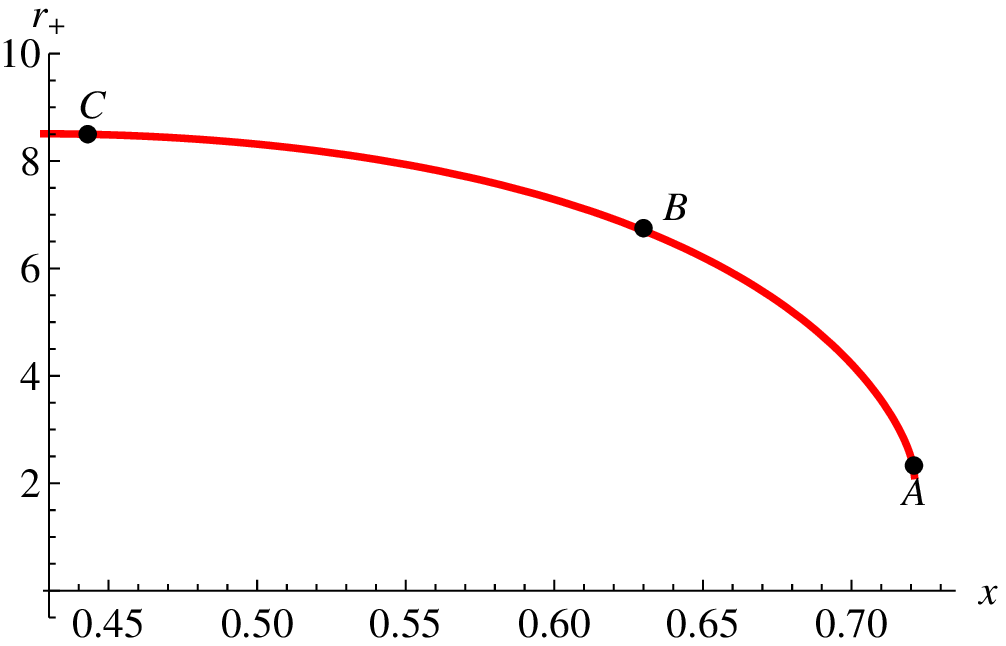}\label{rx25}}~~~
\subfigure[$~~\bar\phi=0.002,~P_{eff}=1.459\times 10^{-4}$]{\includegraphics[width=0.35\textwidth]{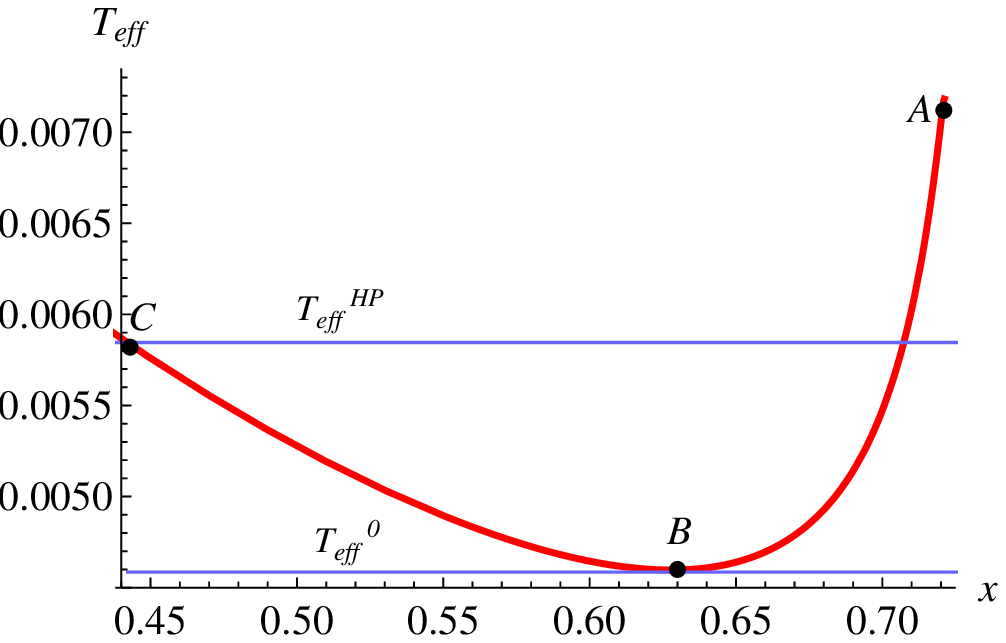}\label{tx25}}~~~
\caption{The pictures of $T_{eff}-x$ and $r_+-x$ for the parameters $k=1,~q=1$. \label{trx25}}
\end{figure}

In order to analyze the different stable phases in dS spacetime with the increasing or the decreasing of $T_{eff}$ for the given effective pressure and non-linear charge correction, we show the pictures of $G_{P_{eff}}-T_{eff}$, $T_{eff}-r_+$, $r_+-x$, and $T_{eff}-x$ with $\bar\phi=0.002,~P_{eff}=1.459\times 10^{-4}$ in Figs. \ref{gtsr25} and \ref{trx25}. The minimum effective temperature and Hawking-Page temperature are denoted by the uppercase letters $B$ and $C$. And $A$ stands for any temperature of the upper branch in $G_{P_{eff}}-T_{eff}$.

From Figs. \ref{srp}, \ref{tr25}, and \ref{trx25}, we know that the range between $C$ and $B$ stands for the bigger black holes which have the positive heat capacity at constant pressure and are thermodynamically stable, whereas the range between $B$ and $A$ stands for the smaller black holes which have the negative heat capacity at constant pressure and are thermodynamically unstable. There exists a minimum effective temperature below which no dS black hole can exist. The only thermal radiation phase characterized by vanishing Gibbs free energy can stable exist when $T_{eff}<T_{eff}^0$. From Fig. \ref{gt25}, we can see that with the increasing of $T_{eff}$ from zero it becomes possible to form a large dS black hole as $T_{eff}>T_{eff}^0$, whose Gibbs free energy is larger than that of the thermal radiation phase. Therefore, these such bigger dS black holes are thermodynamically metastable . Further increasing the effective temperature to $T_{eff}^{HP}$, the radiation phase and black hole are coexist with the vanishing Gibbs free energy, i.e., the HP phase transition emerges. Above this temperature the radiation phase is collapsing into a large dS black hole, which is the most stable phase. The metastable bigger dS black hole that exist for $T_{eff}^0<T_{eff}<T_{eff}^{HP}$ are often neglected. For a stable large dS black hole with continuous decreasing effective temperature it is possible to pass through HP phase transition point and become a metastable phase. This process is just like a supercooled liquid phase of water below its freezing point.

\begin{figure}[htp]
\subfigure[]{\includegraphics[width=0.35\textwidth]{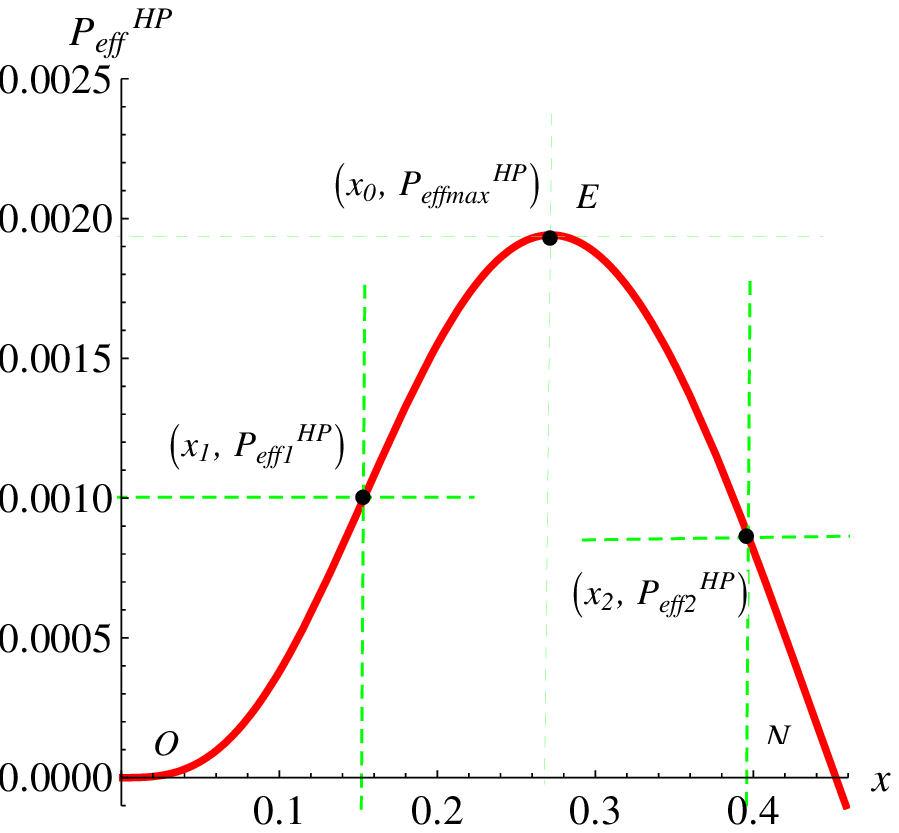}\label{px}}~~~~~~~
\subfigure[]{\includegraphics[width=0.35\textwidth]{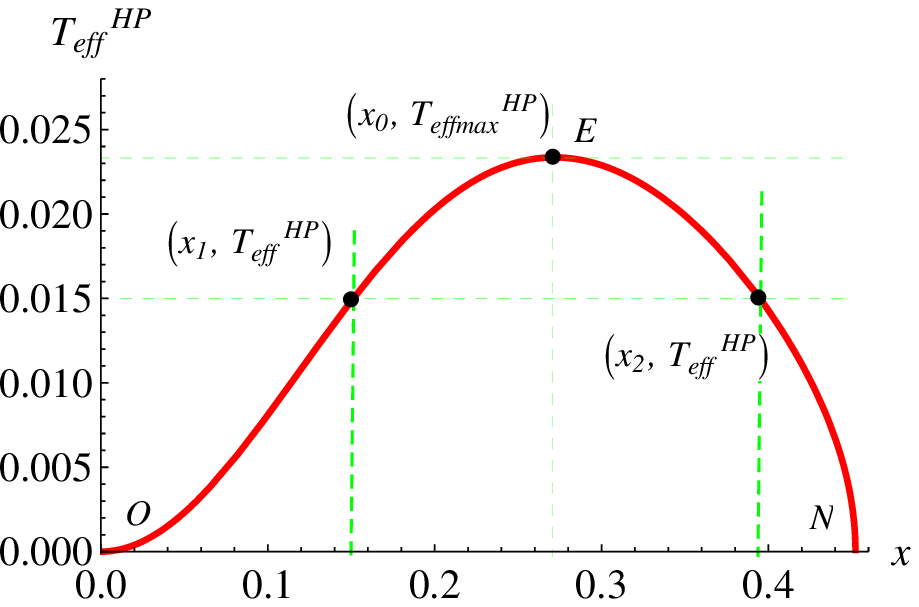}\label{tx}}
\caption{$P_{eff}^{HP}$ and $T_{eff}^{HP}$ as the functions of $x$. The parameters set to $k=1$, $q=1$, and $\bar\phi=0.002$. \label{tpx}}
\end{figure}

At the HP phase transition points, the horizon of dS black hole as a function of $x$ reads
\begin{eqnarray}
r_+=r_G=q\sqrt{\frac{\frac{1-x^4}{2(1-x^3)}+\frac{\pi F(x) f_3(x)}{x^2 f_1(x)}-\frac{4\pi(1-x^3)f_6(x)}{x^2 f_1(x)}}{\frac{\pi F(x) f_2(x)}{x^2 f_1(x)}+\frac{4\pi(1-x^3)f_5(x)}{3x^3 f_4(x)}-\frac{1-x^2}{2(1-x^3)}}}.\label{rg}
\end{eqnarray}
Substitute the above equation into eqs. (\ref{T}) and (\ref{P}), we can obtain the curves of $T_{eff}^{HP}-x$ and $P_{eff}^{HP}-x$ with the given charge and non-linear charge correction in Fig. \ref{tpx}. Unlike in AdS black holes, there exist the upper bound of the HP temperature and HP pressure, i.e., $T_{effmax}^{HP}$ and $P_{effmax}^{HP}$, where are remarked by the uppercase letter $E$. And the ranges of $x$ in diagrams of $T_{eff}^{HP}-x$ and $P_{eff}^{HP}-x$ are both from zero (marked by $O$) to the maximum $x_{max}$ (marked by $N$). The corresponding effective temperature and effective pressure are both zero at $O$ and $E$. Furthermore the values of $x$ at $T_{effmax}^{HP}$ and $P_{effmax}^{HP}$ are the same ($x=x_0$). For any fixed HP temperature lower than $T_{effmax}^{HP}$, there are two values of $x$: $x=x_1$ and $x=x_2$. However the corresponding HP pressures for $x=x_1$ and $x=x_2$ are different: $P_{eff2}^{HP}<P_{eff1}^{HP}<P_{effmax}^{HP}$.

\begin{figure}[htp]
\includegraphics[width=0.35\textwidth]{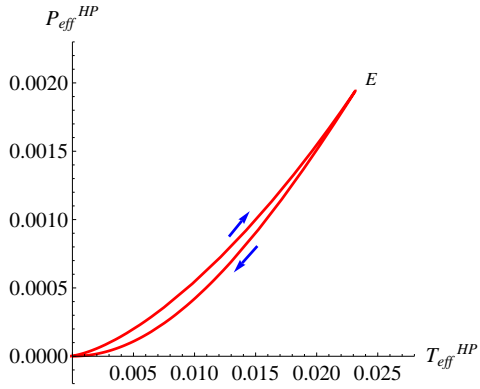}
\caption{$P_{eff}^{HP}$ as the function of $T_{eff}^{HP}$ . The parameters set to $k=1$, $q=1$, and $\bar\phi=0.002$. \label{tp}}
\end{figure}

In order to illustrate this character the coexistence curve of $T_{eff}^{HP}-P_{eff}^{HP}$ has been shown in Fig. \ref{tp}. It is very interesting that the curve of $T_{eff}^{HP}-P_{eff}^{HP}$ is a closed one with two different branches. And the upper branch responds to the process between $O$ and $E$, the lower is that between $N$ and $E$. That means with the increasing ratio $x$ from zero to $x_0$ (i.e., the distance $d=\frac{r_+(1-x)}{x}$ between two horizons decreases from $\infty$ to $\frac{r_+(1-x_0)}{x_0}$), the spacetime with the stable coexistence of dS black hole and radiation phase is going along with $O\rightarrow E$. And the effective temperature and effective pressure are both increasing from zero to the maximum. Continue decreasing the distance between two horizons until to $\frac{(1-x_{max})r_+}{x_{max}}$, the spacetime is going along with $E\rightarrow N$, and the effective temperature and effective pressure are both decreasing from maximum to zero.

\begin{figure}[htp]
\subfigure[~~$\bar\phi=0.002$]{\includegraphics[width=0.35\textwidth]{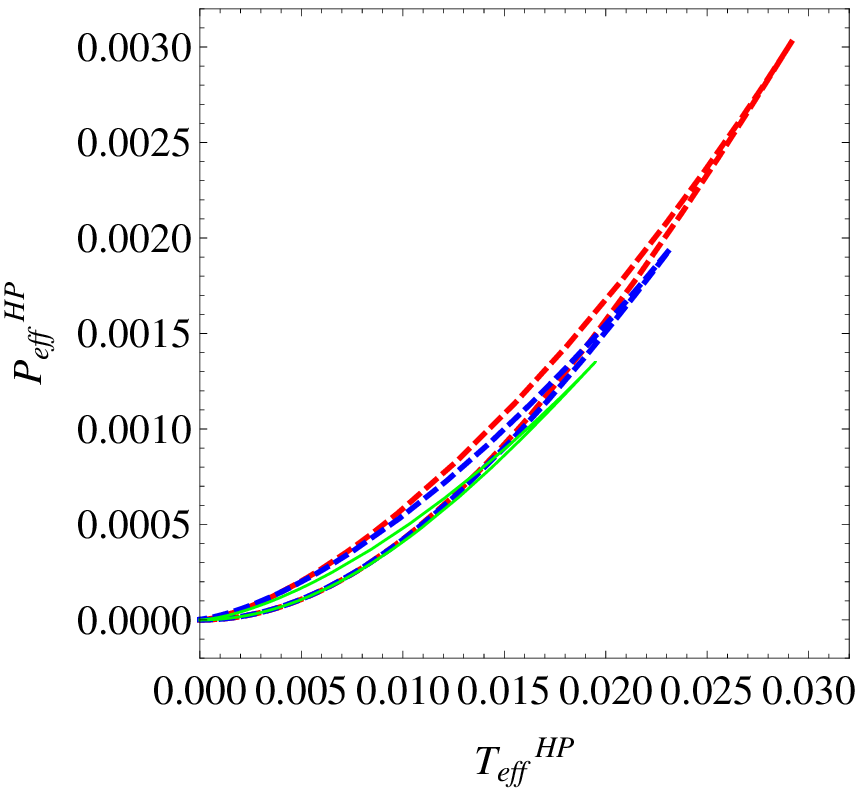}\label{tpq}}~~~
\subfigure[~~$q=1$]{\includegraphics[width=0.35\textwidth]{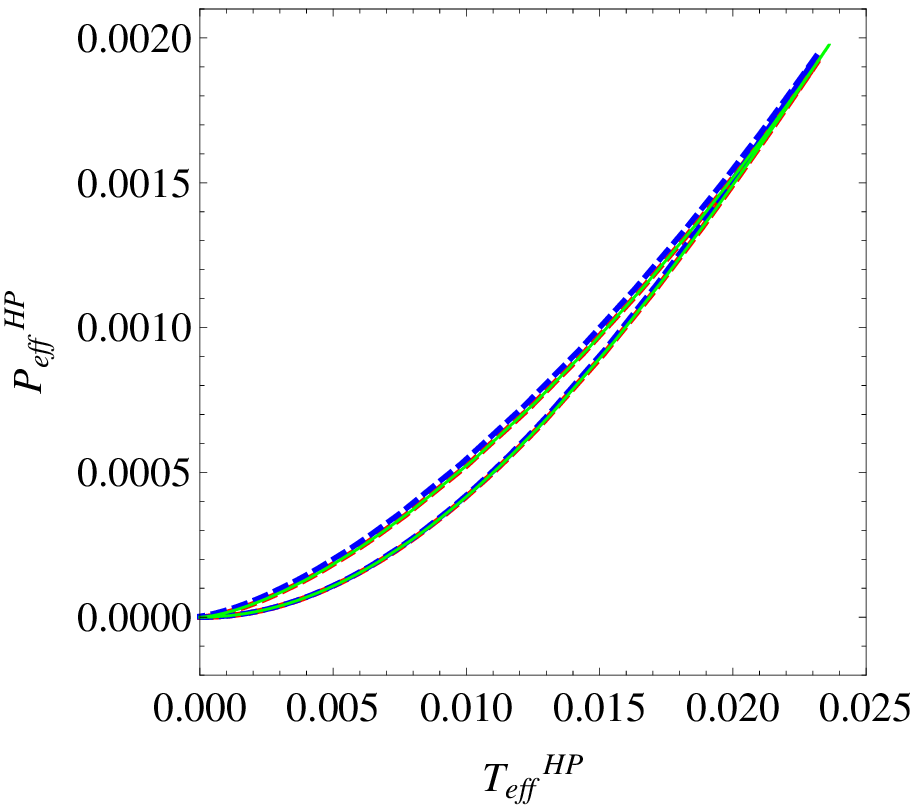}\label{tpphi}}
\caption{The effective pressure $P_{eff}^{HP}$ as functions of $T_{eff}^{HP}$ with different charges and different non-linear charge corrections for $k=1$. In the left the charge set to $q=0.8$ (dashed red thick line), $q=1$ (dashed blue thick line), and $q=1.2$ (green thin line), respectively. In the right the non-linear charge correction set to $\bar\phi=0$ (dashed red thick line), $\bar\phi=0.002$ (dashed blue thick line), and $\bar\phi=0.005$ (green thin line), respectively.
\label{tpqphi}}
\end{figure}

In addition the effects of the charge and non-linear charge correction on the coexistent curve are displayed by Fig. \ref{tpqphi}. We find that for the given non-linear charge correction, $T_{eff}^{HP}$ and $P_{eff}^{HP}$ are both sensitive to the charge, and the maximum of them are both significantly decreasing with the increasing of charge (see the Fig. \ref{tpq}). However for the given charge, the non-linear charge correction has little effect on the coexistence (see the Fig. \ref{tpphi}).

\section{discussions and conclusions}
\label{four}

In this paper, we mainly have analyzed the HP phase transition of the four-dimensional topological dS spacetime with the non-linear charge correction, which can be regarded an ordinary thermodynamic system in the thermodynamic equilibrium.

Firstly we reviewed the thermodynamical quantities of the four-dimensional dS spacetime with the non-linear charge correction and in isobaric processes given the corresponding pictures of the effective temperature, entropy, and the dS black hole horizon with the certain range of $x$. We found in an isobaric process the effective temperature is not a monotonously function with $r_+$, and the smaller dS black holes are of the negative heat capacity and the bigger ones are with $C_{P_{eff}}>0$. That means in dS spacetime with $T_{eff}>T_{eff}^0$ there exist the stable bigger dS black holes, instead of the smaller dS black holes. Furthermore there exists the minimal effective temperature $T_{eff}^0$ for the dS black hole with fixed effective pressure and non-linear charge correction. As $T_{eff}<T_{eff}^0$, no dS black hole can survive. While a pair of dS black holes emerge when $T_{eff}>T_{eff}^0$.

Then we investigated the thermodynamic property of HP phase transition with different effective pressure and different non-linear charge correction. We found that two branches are in $G_{P_{eff}}-T_{eff}$, the upper one stands for the unstable smaller dS black holes and the lower one is the stable bigger dS black holes. Furthermore there both exist the minimum effective temperature and the HP temperature, $T_{eff}^0<T_{eff}^{HP}$. While compared with the pure thermal radiation phase characterized by vanishing Gibbs free energy, the dS spacetime is of the thermal radiation phase, not of the bigger black hole. As $T_{eff}=T_{eff}^{HP}$, the radiation phase and black hole are coexisted with the vanishing Gibbs free energy, i.e., the HP phase transition emerges. Above this temperature the thermal radiation phase is collapsing into a large dS black hole, which is the most stable phase. In addition the Gibbs free energy is sensitive to the effective pressure, not the non-linear charge correction.

Finally we presented the coexistence curve of $T_{eff}^{HP}-P_{eff}^{HP}$ in Fig. \ref{tp}. It is very interesting and unique that the coexistent curve is a closed one with two different branches. And the HP temperature and HP pressure both are bounded from zero to the maximum. That is fully different from the AdS black hole, whose temperature and pressure at HP phase transition point are both from zero to infinity. We can regard the existent curve of HP phase transition as a difference between dS spacetime and AdS black hole. Furthermore with the decreasing of the distance between two horizons from $\infty$ to $\frac{(1-x_0)r_+}{x_0}$, the spacetime with the coexistent state of dS black hole and thermal radiation phase is going along with the above branch. And the effective temperature and effective pressure are both increasing from zero to the maximum. Continue decreasing the distance between two horizons until to $\frac{(1-x_{max})r_+}{x_{max}}$, the spacetime is going along with the lower branch, and the effective temperature and effective pressure are both decreasing from maximum to zero.

\section*{Acknowledgments}
We would like to thank Prof. Ren Zhao and Meng-Sen Ma for their indispensable discussions and comments. This work was supported by the Natural Science Foundation of China (Grant No. 11705106, Grant No. 11475108, Grant No. 12075143), the Scientific Innovation Foundation of the Higher Education Institutions of Shanxi Province (Grant Nos. 2020L0471, Grant Nos. 2020L0472), and the Science Technology Plan Project of Datong City, China (Grant Nos. 2020153).

\end{document}